\documentclass[a4paper,11pt]{article}

\usepackage{jcappub} 

\usepackage[T1]{fontenc} 
\usepackage{xcolor}
\usepackage{bm}
\usepackage{appendix}
\usepackage{enumerate}

\allowdisplaybreaks[4]

\title{\boldmath Probing the vector charge of Sagittarius A* with pulsar timing}

\author[a,b]{Zexin Hu,}
\author[b,d,1]{Lijing Shao,\note{Corresponding author.}}
\author[c]{Rui Xu,}
\author[b]{Dicong Liang,}
\author[b]{and Zhan-Feng Mai}

\affiliation[a]{Department of Astronomy, School of Physics, Peking University, Beijing 100871, China}
\affiliation[b]{Kavli Institute for Astronomy and Astrophysics, Peking University, Beijing 100871, China}
\affiliation[c]{Department of Astronomy, Tsinghua University, Beijing 100084, China}
\affiliation[d]{National Astronomical Observatories, Chinese Academy of Sciences, Beijing 100012, China}

\emailAdd{huzexin@pku.edu.cn}
\emailAdd{lshao@pku.edu.cn}
\emailAdd{xuru@tsinghua.edu.cn}
\emailAdd{dcliang@pku.edu.cn}
\emailAdd{zhanfeng.mai@gmail.com}

\abstract{Timing a pulsar orbiting around Sagittarius~A* (Sgr~A*) can provide us
with a unique opportunity of testing gravity theories. We investigate the
detectability of a vector charge carried by the Sgr~A* black hole (BH) in the
bumblebee gravity model with simulated future pulsar timing observations.  The
spacetime of a bumblebee BH introduces characteristic changes to the orbital
dynamics of the pulsar and the light propagation of radio signals.  Assuming a
timing precision of 1\,ms, our simulation shows that a 5-yr observation of a
pulsar with an orbital period $P_b\sim 0.5\,{\rm yr}$ and an orbital
eccentricity $e\sim 0.8$ can probe a vector charge-to-mass ratio as small as
$Q/M\sim 10^{-3}$, which is much more stringent than the current constraint from
the Event Horizon Telescope (EHT) observations, and comparable to the
prospective constraint from extreme mass-ratio inspirals with the Laser
Interferometer Space Antenna (LISA).}

\begin{document}
\maketitle
\flushbottom

\section{Introduction}
\label{sec:intro}

Due to the extreme rotation stability, pulsars are regarded as one of the best
probes to test gravity theories~\cite{Damour:2007uf, Kramer:2004hd,
Shao:2022izp, Hu:2023vsq}. With the observational data of 16 years, the Double
Pulsar system provides the most precise test of general relativity (GR) in a
system containing strongly self-gravitating bodies~\cite{Kramer:2021jcw}. It
also has been anticipated that timing a pulsar orbiting closely around the
supermassive black hole (SMBH) residing in our Galactic Center (GC), the
so-called Sagittarius A* (Sgr~A*), can provide us with a unique opportunity of
studying the black hole (BH) physics as well as the environment around the
GC~\cite{Liu:2011ae, Shao:2014wja, Psaltis:2015uza, Zhang:2017qbb,
Bower:2018mta, Weltman:2018zrl, 2019BAAS...51c.438B, EHT:2019nmr, Dong:2022zvh, 
EventHorizonTelescope:2022xqj, Hu:2023ubk}.

From both theoretical models and observational evidences of the stellar
population around Sgr A*, we expect the existence of a large number of neutron
stars in the GC~\cite{Paumard:2006im,Lu:2013sn,Wharton:2011dv}. The discovery of
the magnetar in this region also strongly suggests the existence of normal
pulsars as magnetars are thought to be rare pulsars~\cite{Eatough:2013nva}.
However, due to the large dispersion measures and highly turbulent interstellar 
medium in the GC~\cite{Abbate:2023car}, no normal pulsar has been found yet
within the inner parsec even in those searches performed at high radio
frequencies~\cite{Kramer:2000tc,Siemion:2013,Liu:2021ziv, EHT:2023hcj}. 
Nevertheless, future observations with the next-generation facilities, such as
the Square Kilometre Array (SKA) and the next-generation Very Large Array
(ngVLA), are expected to find a number of pulsars in this region. We also expect
to time these pulsars well, which will open a new avenue of testing gravity
theories~\cite{Liu:2011ae, Shao:2014wja, Bower:2018mta, Shao:2020fka,
Dong:2022zvh, Eatough:2023tst}. 

In this work we focus on the possibility of using a pulsar-SMBH system to
constrain the vector charge carried by the Sgr~A* in a specific non-minimally
coupled vector-tensor theory, the so-called bumblebee gravity model. The action
of this  theory reads~\cite{Kostelecky:2003fs}
\begin{equation}\label{eq:action}
  S=\int {\rm d}^4 x
  \sqrt{-g}\left(\frac{1}{2\kappa}R+\frac{\xi}{2\kappa}B^{\mu}B^{\nu}R_{\mu
  \nu}-\frac{1}{4}B^{\mu\nu}B_{\mu\nu}-V\right)+S_m\,,
\end{equation}
where $S_m$ is the action for conventional matter, $\kappa=8\pi G$, $B_{\mu\nu}
\equiv D_{\mu}B_\nu-D_\nu B_\mu$ is the field strength of the vector field
$B_{\mu}$ with $D_{\mu}$ being the covariant derivative, $\xi$ is the coupling
constant of the non-minimal coupling between the bumblebee vector and the Ricci
tensor $R_{\mu\nu}$, and $V$ is the potential for the vector field $B_\mu$. The
bumblebee gravity model has been widely studied in the literature as an example
for spontaneous violation of the Lorentz symmetry or for being a simple
extension of the Einstein-Maxwell theory~\cite{Kostelecky:2003fs, Bluhm:2004ep,
Liang:2022hxd, Xu:2022frb}. While observations in the Solar System set stringent
constraints on the bumblebee gravity in the weak-field
region~\cite{Hellings:1973zz,Xu:2022frb}, recent studies using the BH shadows
observed by the Event Horizon Telescope (EHT) leave a large parameter space in
the strong-field region untested~\cite{Xu:2022frb,Xu:2023xqh}. Considering the
high precision of pulsar timing experiments, the main purpose of this work is to
develop a timing model for pulsar-SMBH systems in the bumblebee gravity theory
and to investigate the potential of constraining the bumblebee gravity with
future timing observations.

The paper is organized as follows. In Sec.~\ref{sec:BBBH}, we briefly overview
the numerical BH solutions in the bumblebee gravity theory and derive a leading
order solution expanded in terms of the bumblebee charge of the BH. We construct
two pulsar timing models in Sec.~\ref{sec:timing}, which include calculations
for the beyond-GR light propagation and orbital motion in the BH
spacetime. The simulation results of possible constraints on the bumblebee
charge of Sgr~A* are shown in Sec.~\ref{sec:esti} and we have some final
discussion in Sec.~\ref{sec:discussion}.

In the remaining part of this paper, we use the geometrized unit system where
$G=c=1$, except when traditional units are written out explicitly. The sign
convention of the metric is $(-,+,+,+)$.

\section{Bumblebee BHs}
\label{sec:BBBH}

The field equations in the bumblebee gravity are obtained by taking variations
of the metric field $g_{\mu\nu}$ and the bumblebee field $B^{\mu}$ in
Eq.~(\ref{eq:action}). Detailed calculation can be found in
Ref.~\cite{Xu:2022frb}. For convenience, we briefly review the construction of
the static spherical BH solutions here. To focus on the static spherical BH 
solutions, we adopt the metric ansatz
\begin{equation}\label{eq:spherical ansatz}
  {\rm d}s^2=-e^{2\nu}{\rm d}t^2+e^{2\mu}{\rm d}r^2+r^2\left({\rm
  d}\theta^2+\sin^2\theta{\rm d} \phi^2\right)\,,
\end{equation}
as well as the assumption that $B_{\mu}=(b_t,0,0,0)$, where $\nu$, $\mu$, and
$b_t$ are functions of $r$ only.  Following the assumptions in
Ref.~\cite{Xu:2022frb}, we ignore the contribution from the potential $V$ which,
in the setting of spontaneous Lorentz symmetry breaking, merely plays a role of
vacuum energy after giving the background field values of $B_\mu$. Then the
vacuum field equations can be simplified to ordinary differential equations for
$\nu$, $\mu$, and $b_t$~\cite{Xu:2022frb,Liang:2022gdk},
\begin{eqnarray}
  0&=&e^{2\mu}-2r\nu'-1-r^2 e^{-2\nu}\left(\frac{\kappa}{2}b_t'^2-\xi b_t
  b_t'\nu'+\xi b_t^2 \nu'^2\right)\,,\label{eq:field eqs1}\\
  0&=&\mu'\nu'+\frac{1}{r}\mu'-\frac{1}{r}\nu'-\nu''-\nu'^2+e^{-2\nu}\left(\frac{\kappa}{2}
  b_t'^2-\xi b_t b_t'\nu'+\xi b_t^2\nu'^2\right)\,,\label{eq:field eqs2}\\
  0&=&b_t''-b_t'\left(\mu'+\nu'-\frac{2}{r}\right)+\frac{\xi}{\kappa}b_t\left(\mu'\nu'-\frac{2}
  {r}\nu'-\nu''-\nu'^2\right)\,,\label{eq:field eqs3}
\end{eqnarray}
where the prime denotes the derivative with respect to $r$.

The general assumption for $B_\mu$ with spherical symmetry should contain a
radial component $b_r$. However, as discussed by \citet{Xu:2022frb}, the BH
solutions with a vanishing $b_r$ can recover the Reissner-Nordstr\"om (RN)
solution when $\xi$ tends to zero, while the solutions with a nonzero $b_r$ only
exist when $\xi\neq 0$. It means that the solutions with a nontrivial $b_r$ are
extraordinary solutions caused by the nonminimal coupling~\cite{Liang:2022gdk}.
For simplicity, we only consider the ordinary solutions that are connected to
the RN solution in the present work.

\subsection{Numerical solutions}

A general analytical solution of Eqs.~(\ref{eq:field eqs1}--\ref{eq:field eqs3})
has not been found yet, thus we can only obtain the BH solutions numerically.
Following Refs.~\cite{Xu:2022frb,Liang:2022gdk}, one first eliminates $\mu$ with
Eq.~(\ref{eq:field eqs1}) to obtain two second-order ordinary differential
equations for $\nu$ and $b_t$.  Then one can integrate the system outward from
the horizon $r=r_h$. The initial conditions that ensure a BH solution have been
carefully discussed in Ref.~\cite{Xu:2022frb}, which read
\begin{eqnarray}
  e^{2\nu}&=&N_{11}(r-r_h)+N_{12}(r-r_h)^2+\cdots\,,\\
  b_t&=&L_{11}(r-r_h)+L_{12}(r-r_h)^2+\cdots\,,
\end{eqnarray}
where $N_{11},L_{11}$ are free parameters and
\begin{eqnarray}
  N_{12}&=&-\frac{N_{11}}{r_h}+\left(\frac{\kappa}{2}-\frac{\xi}{4}\right)L_{11}^2+\left(\frac
  {\kappa\xi}{4}-\frac{\xi^2}{4}+\frac{\xi^3}{16\kappa}\right)\frac{L_{11}^4
  r_h}{N_{11}}\,,\\
  L_{12}&=&-\frac{L_{11}}{r_h}+\left(\frac{\xi}{4}-\frac{\xi^2}{8\kappa}\right)\frac{L_{11}^3}
  {N_{11}}+\left(\frac{\kappa\xi}{4}-\frac{\xi^2}{4}+\frac{\xi^3}{16\kappa}\right)\frac{
  L_{11}^5 r_h}{N_{11}^2}\,.
\end{eqnarray}

After integrating outward with given $r_h$, $N_{11}$, and $L_{11}$, one needs to
redefine the time coordinate such that $g_{tt}=-e^{2\nu}$ goes to $ -1$ as
$r\rightarrow\infty$. Thus only two of these three parameters are independent,
which are equivalent to the mass $M$ and bumblebee charge $Q$ of the BH that are
defined with the asymptotic behavior of $\nu$ and $b_t$ at spacial 
infinity~\cite{Xu:2022frb,Liang:2022gdk}
\begin{align}
  M &\equiv-\lim_{r\rightarrow\infty}r^2\mu'=\lim_{r\rightarrow\infty}r^2\nu'\,,\\ 
  Q &\equiv-\sqrt{\frac{\kappa}{2}}\lim_{r\rightarrow\infty}r^2b_t'\,.
\end{align}
The factor $\sqrt{\kappa/2}$ in the definition of $Q$ is chosen so that it can
recover the RN solution when $\xi=0$. With this factor, $Q$ then can be
interpreted as an extension of the electric charge. The bumblebee charge $Q$ can
be positive or negative depending on the sign of $b_t$. As one can see from
Eqs.~(\ref{eq:field eqs1}--\ref{eq:field eqs2}) that the metric functions only
depend on $b_t$ in quadratic form, we only consider $Q\geq0$ in this paper.

There are two special cases for the BH solutions in the bumblebee
gravity~\cite{Xu:2022frb, Liang:2022gdk}. One is $\xi=0$, in which the theory
reduces to the Einstein-Maxwell theory so that the BH solution recovers the RN
solution.  The other case is $\xi=2\kappa$. In this configuration, the BH
solution is described by a Schwarzchild metric accompanied by a nonvanishing
bumblebee field. Similar solutions exist in many other vector-tensor theories
and scalar-tensor theories~\cite{Heisenberg:2017hwb,
Cisterna:2016nwq,Babichev:2017guv}. In this case, one cannot distinguish the
charged BH in the bumblebee theory from a Schwarzchild BH in GR with only the
background BH metric considered~\cite{Barausse:2008xv}.

\subsection{Leading order solutions in $Q/M$}
\label{subsec:analytic metric}

Although current observations with EHT and thermodynamic consideration of the
bumblebee BHs  only set loose constraints on $Q/M$ for Sgr~A* even for the case
that $\xi$ deviates from $2\kappa$
significantly~\cite{Xu:2022frb,Xu:2023xqh,Mai:2023ggs}, it turns out that the
prospective measurement precision of $Q/M$ using pulsar timing can be as small
as $Q/M\sim \mathcal{O}(10^{-2})$ for $| \xi/2\kappa-1|\gtrsim 1$ as we will
find out later.  Moreover, from Eqs.~(\ref{eq:field eqs1}--\ref{eq:field eqs2}) 
one can see that the influence of the bumblebee field $b_t$ only appears in
quadratic form, which means that the next-to-leading order effects caused by the
bumblebee charge will be $ \mathcal{O}(10^{-4})$ smaller than the leading order
effects in the cases that we are interested in. Thus a BH solution accurate to
the leading order of $Q/M$ can be used in constructing our timing model for a
pulsar around  Sgr A* which is equipped with a bumblebee charge. Here we present
such an analytic approximate solution as follows
\begin{eqnarray}
  b_t &=& -\frac{1}{\sqrt{2\kappa}}\left(1-\frac{2M}{r}\right)\frac{Q}{M}+
  \mathcal{O}\left(\frac{Q}{M} \right)^3\,,\\
  g_{tt} &=& -1+\frac{2M}{r}-\left(1-\frac{\xi}{2\kappa}\right)\frac{Q^2}{r^2}+
  \mathcal{O}\left(\frac{Q} {M}\right)^4\,,\\
  g_{rr} &=&
  \left[1-\frac{2M}{r}+\left(1-\frac{\xi}{2\kappa}\right)\frac{Q^2}{r^2}\right]^{-1}
  +\mathcal{O}\left(\frac{Q}{M}\right)^4\, ,
\end{eqnarray}
and the detailed derivation can be found in Appendix~\ref{app:order solution}.
The radius of the event horizon $r_h$ can  also  be determined to the same
order, 
\begin{equation}
  \frac{r_h}{M}=1+\sqrt{1-\left(1-\frac{\xi}{2\kappa}\right)\left(\frac{Q}{M}\right)^2}+
  \mathcal{O}\left(\frac{Q}{M}\right)^4\,,
\end{equation}
where the retainment of the square root shows the recovery of the RN solution
for $\xi=0$ explicitly.

An important observation from the leading order solution is that, in the leading
order metric functions, $Q$ and $\xi$ always appear as a combined factor
$(1-\xi/2\kappa)Q^2$, which means that we cannot distinguish a charged BH in the
bumblebee theory from a RN BH in GR when $Q$ is small and $\xi/2\kappa<1$ by
measuring the background metric only. Such a degeneracy can be broken by 
considering higher order terms as shown in Appendix~\ref{app:order solution}, or
probably by metric perturbations.

\section{Pulsar timing model}
\label{sec:timing}

As studied in Refs.~\cite{Liu:2011ae, Psaltis:2015uza, Zhang:2017qbb, Hu:2023ubk}, timing a
pulsar in a close orbit around Sgr~A* with future instruments like SKA or ngVLA
can provide us with a unique opportunity to measure BH properties to a high 
precision.  
A generic numerical model of pulsar-SMBH systems was developed in Ref.~\cite{Hu:2023ubk}.
To consider the possibility of constraining the bumblebee charge of
Sgr~A* with pulsar timing, the first step is to construct a timing model that
consistently includes the effects caused by the bumblebee charge.  In pulsar
timing, the timing model connects a pulsar's proper rotation with the observed
times of arrival (TOAs) at radio telescopes by modeling various physical effects
taking part in the travel history of the radio signal from the pulsar to the
Earth~\cite{Damour:1986, Damour:1991rd, Hu:2023vsq}.  Ignoring the aberration
effect~\cite{Damour:1986}, the pulse arrival time $t^{\rm TOA}$ in the 
barycenter of the Solar System is related to the pulsar's proper time $T$ via
\begin{equation}\label{eq:T to TOA}
  t^{\rm TOA}=T+\Delta_{\rm E}+\Delta_{\rm prop}\,,
\end{equation}
where the Einstein delay $\Delta_{\rm E}=t-T$ translates the proper time of the
pulsar to the coordinate time $t$, and $\Delta_{\rm prop}$ accounts for the
propagation effects. The rotation of the pulsar is usually described by a simple
rotation model~\cite{Lorimer:2005misc}
\begin{equation}
  N=N_0+\nu_p T+\frac{1}{2}\dot{\nu}_p T^2+\cdots\,,
\end{equation}
where $N$ is the rotation number and $\nu_p$,  $\dot{\nu}_p$ are the pulsar's
spin and spin down rate respectively. Combining these two equations, one can
calculate the TOA for each pulse after knowing the propagation of light and
pulsar's orbital motion.

In Sec.~\ref{subsec:light} and Sec.~\ref{subsec:orbit}, we develop the timing
model that fully uses a numerical method to account for all the possible effects
caused by the bumblebee charge of the BH. We also provide a timing model in 
Sec.~\ref{subsec:analytic model} based on the leading order solution introduced
above.

\subsection{Light propagation}
\label{subsec:light}

The propagation time delay $\Delta_{\rm prop}$ in Eq.~(\ref{eq:T to TOA})
represents the time delay during the pulse signal travelling from the pulsar to
the Earth. Here we focus on the additional time delay caused by the bumblebee
charge and ignore the various effects caused by the interstellar medium, 
Galactic acceleration, as well as the translation between the Solar System
barycenter and the local coordinate time on the Earth~\cite{Damour:1991rd,
Lorimer:2005misc}.  The distance between Sgr~A* and the Solar System is treated
as infinity, which means we also ignore the proper motion of
Sgr~A*~\cite{Reid:2004rd}. These simplifications are adopted as we are only
aiming to estimate the measurability of the bumblebee charge carried by the
Sgr~A*.  While for timing models used in real observations one should take into
account all these effects depending on the real timing precision.

In the spherical spacetime represented by Eq.~(\ref{eq:spherical ansatz}), the
light-like geodesic equation can be rewritten with the help of its first
integrals as
\begin{align}\label{eq:light prop}
  \left(\frac{{\rm d}r}{{\rm d}t}\right)^2 &= e^{2\nu-2\mu}\left(1-\frac{b^2
  e^{2\nu}}{r^2} \right)\,,\\ 
  \frac{{\rm d}\phi}{{\rm d}t} &=\frac{be^{2\nu}}{r^2}\,, \label{eq:light prop2}
\end{align}
where $b$ is the impact parameter and we have set the orbit in the equatorial
plane because of the spherical symmetry. For a pulsar at a given position,
solving the light propagation time is the so-called emitter-observer problem,
which in general dose not have an analytic solution even in the spherical
spacetime~\cite{Klioner:2010pu, DellaMonica:2023ydm}. With the metric solved 
numerically in Sec.~\ref{sec:BBBH}, one can first solve the impact parameter $b$
with shooting method from the equation
\begin{equation}\label{eq:eq of b}
  \Delta\phi=\int_{r\rightarrow\infty}\frac{{\rm d}\phi}{{\rm d}r}{\rm d}r\,,
\end{equation}
where $\Delta\phi$ is the angle between the pulsar and the Earth as seen from
Sgr~A*. The range of $\Delta\phi$ is considered to be between $0$ and $\pi$,
which corresponds to the direct propagation paths. The light paths with
$\Delta\phi>\pi$ bend so strongly such that they  in general result in images
too faint to be observed~\cite{Lai:2004ue}. One should also notice that the
integration in Eq.~(\ref{eq:eq of b}) splits into two pieces with ${\rm d}r/{\rm
d}\phi$ taking different signs if the path has a periastron at ${\rm d}r/{\rm
d}\phi=0$.

\begin{figure}[tbp]
\centering 
\includegraphics[width=14cm]{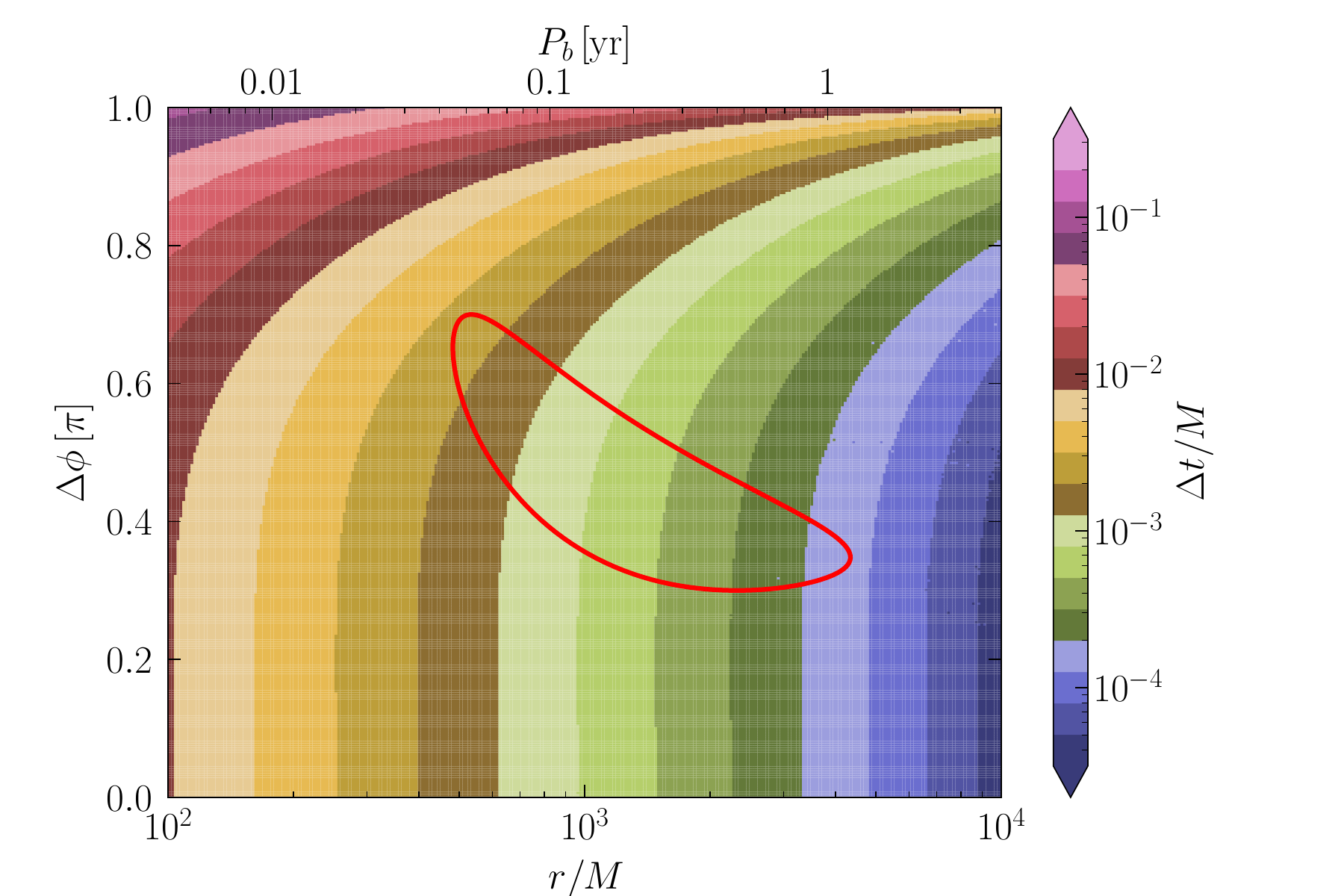}
\caption{\label{fig:prop time diff} The difference in the light propagation time
$\Delta t= \big|\Delta_{\rm prop}^{\rm Bum}-\Delta_{\rm prop}^{\rm GR}\big|$
between GR and the bumblebee gravity for $\xi=\kappa$ and $Q=M$. The bottom
horizontal axis $r$ is the distance between the pulsar and Sgr~A*, and the top
horizontal axis $P_b$ is the corresponding period of the pulsar on a circular
orbit.  The vertical axis $\Delta\phi$ is the angle between the pulsar and the
Earth as seen from Sgr~A*.  The red line corresponds to an eccentric Newtonian orbit with
orbital parameters shown in Eq.~(\ref{eq:orbit params}).}
\end{figure}

One can directly integrate Eqs.~(\ref{eq:light prop}--\ref{eq:light prop2}) to
obtain the light propagation time $\Delta_{\rm prop}$ after getting the impact
parameter $b$. In Fig.~\ref{fig:prop time diff}, we plot the time difference
$\Delta t=\big|\Delta_{\rm prop}^{\rm Bum}-\Delta_{\rm prop}^{\rm GR} \big|$ 
between GR and the bumblebee gravity for $\xi=\kappa$ and $Q=M$ as an example.
The red line in the figure illustrates a pulsar orbit with an orbital period
$P_b=0.5\,{\rm yr}$ and an orbital eccentricity $e=0.8$.  Although this example
is somewhat unrealistic for its extremely large $Q$, we illustrate it in order
to show the deviation from GR clearly.  For a pulsar with orbital period
$P_b\sim 0.1\,{\rm yr}$ around Sgr~A*, which has a mass $M\sim 4.3\times
10^6\,M_\odot$, the typical time difference is $\sim20\,{\rm ms}$, which may be
detectable in future observations as the expected timing precision for such a
pulsar can reach $0.1$--$1$ ms~\cite{Liu:2011ae}.  The time difference can be
even larger if the pulsar is in an orbit with its inclination angle close to
$\pi/2$ or has a large orbital eccentricity. However, this time difference is
always proportional $(Q/M)^2$ at the leading order.  For $Q/M\lesssim0.1$, the
time difference is about or less than $1\,{\rm ms}$ for pulsars in the 
interested parameter space, which are hard to be directly detected in the near
future. Nevertheless, one still should at least include its leading order effect
in the timing model as the residuals caused by this effect might be close to the
timing precision, similar to the spirit in the treatment of those 
next-to-leading order effects in the Double Pulsar~\cite{Kramer:2021jcw}.

\subsection{Orbital motion}
\label{subsec:orbit}

As the mass ratio of the pulsar and the Sgr~A* BH is smaller than $10^{-6}$, we
will treat the pulsar as a test particle in the BH's spacetime for the moment.
With this assumption, we also neglect the gravitational radiation that causes
the decay of orbital period, usually encoded in the $\dot{P}_b$ parameter. 
Therefore, the equations of motion for the pulsar read as
\begin{align}\label{eq:orbit}
  \left(\frac{{\rm d}r}{{\rm d}t}\right)^2 &=
  e^{2\nu-2\mu}\left(1-\frac{l^2}{\epsilon^2}
  \frac{e^{2\nu}}{r^2}-\frac{e^{2\nu}}{\epsilon^2}\right)\,,\\ 
  \frac{{\rm d}\phi}{{\rm d}t} &=\frac{l}{\epsilon}\frac{e^{2\nu}}{r^2}\,,\ \
  \frac{{\rm d}T}{{\rm d}t}=\frac{1}{\epsilon} e^{2\nu}\,, \label{eq:orbit2}
\end{align}
where $\epsilon$ is the specific energy constant, and $l$ is the specific
angular momentum constant. By numerically integrating the above equations, one
can obtain the orbital motion of the pulsar, as well as the Einstein delay
$\Delta_{\rm E}$.

As a fiducial case, we study a pulsar with the following orbital parameters,
\begin{eqnarray}\label{eq:orbit params}
  P_b=0.5\,{\rm yr}\,,\quad\quad 
  e=0.8\,,\quad \quad
  i=\frac{1}{5}\pi\,,\quad \quad
  \omega=\frac{5}{7}\pi\,,\quad \quad
  \sigma=-\frac{1}{3}\pi\,,
\end{eqnarray}
where $P_b$ is the orbital period, $e$ is the orbital eccentricity, $i$ is the
inclination angle, $\omega$ is the longitude of the periastron at $t=0$, and
$\sigma$ is the initial orbital phase of the pulsar. As we have ignored the
proper motion of Sgr~A*, the longitude of the ascending node, $\Omega$, will not
affect the TOAs and we choose to fix $\Omega=0$~\cite{Taylor:1994zz}. The 
orbital parameters of the pulsar is understood in a Keplerian manner at $t=0$.
For example, the semi-major axis  $a$ of the orbit is defined through
$a=(MP_b^2/4\pi^2)^{1/3}$, and the pericenter and apocenter distances are given
by $r_{\rm min}=a(1-e)$ and $r_{\rm max}=a(1+e)$ respectively. The relation 
between the parameters $(\epsilon,l)$ and $(P_b,e)$ is determined by the
assumption that $r_{\rm min}$ and $r_{\rm max}$ are the two roots that result in
${\rm d}r/{\rm d}t=0$ in Eqs.~(\ref{eq:orbit}--\ref{eq:orbit2}).

\begin{figure}[tbp]
  \centering 
  \includegraphics[width=14cm]{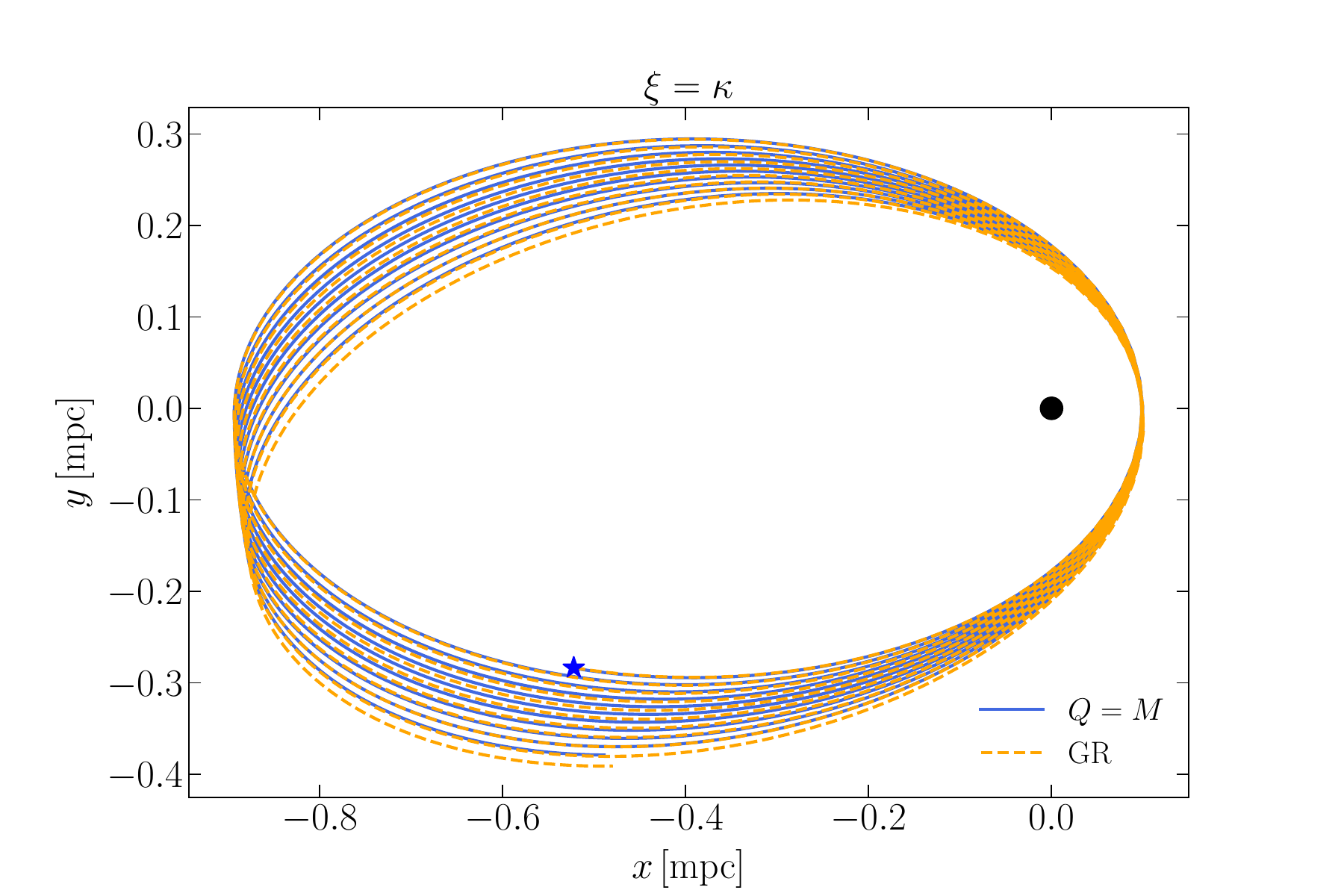}
  \caption{\label{fig:orbit} Five-year pulsar orbits in GR (orange dashed line)
  and in the bumblebee gravity with $\xi=\kappa$ and $Q=M$ (blue solid line).
  The pulsar's orbital parameters are given in Eq.~(\ref{eq:orbit params}). The
  black dot denotes the position of Sgr~A* and the blue star denotes the initial
  location of the pulsar.}
\end{figure}

Due to the assumed spherical symmetry of the BH spacetime, the inclination angle
$i$ is a constant and the motion of the pulsar is in a fixed plane.
Figure~\ref{fig:orbit} gives an illustration of the pulsar's orbits in the 
orbital plane for GR and the bumblebee gravity with $\xi=\kappa$ and $Q=M$. The
total time span $t_{\rm total}=5\,{\rm yr}$ is used, and one can see a secularly
increasing deviation between these two orbits.  This secular effect corresponds
to an additional periastron advance caused by the bumblebee charge.  At the
leading order of $Q/M$, it reads~\cite{Xu:2022frb}
\begin{equation}
  k^{Q}=-\frac{1}{6}\left(1-\frac{\xi}{2\kappa}\right)\left(\frac{Q}{M}\right)^2k^{\rm
  1PN}\,,
\end{equation}
where $k^{\rm 1PN}=3M/a(1-e^2)$ is the periastron advance in GR at the first
post-Newtonian (PN) level~\cite{Damour:1986}. The leading order effect caused by
the bumblebee charge is at the 1\,PN level as expected in the parameterized PN
(PPN) framework where the PPN parameter $\beta\neq 1$ for a charged BH in the
bumblebee gravity~\cite{Xu:2022frb}. Compared to its GR counterpart $k^{\rm
1PN}$, $k^{Q}$ is suppressed by a factor of $(Q/M)^2$ and thus becomes much 
smaller when $Q/M\lesssim0.1$. However, depending on the value of $Q/M$, this
additional periastron advance might be still comparable to the effects caused by
the spin of Sgr~A*, which is numerically at the 1.5\,PN
level~\cite{Wex:1998wt,Liu:2011ae}. Thus it is important in the timing model.
Besides the secular effect, we also include all the periodic effects caused by
the bumblebee charge by resorting to numerical calculations, although they are
in general smaller than the leading order secular effect. 

\subsection{Leading order effects in $Q/M$}
\label{subsec:analytic model}

Combining the results of orbital motion and light propagation, one obtains the
numerical timing model for the pulsar-SMBH system in the bumblebee gravity.
However, as we discussed before, for $Q/M\lesssim0.1$, it might be unnecessary
to calculate all these effects numerically. A leading order approximation should
work for the purpose of forecasting the measurability of the bumblebee charge
while higher order corrections can be added in real timing models depending on 
the timing accuracy of the real data.

To construct a timing model with leading order effects in $Q/M$, we adopt the
expansion in $Q/M$ as well as the usual PN expansion~\cite{Damour:1985}, which
describes the pulsar's motion in the harmonic coordinate. The relation between
the radial position $R$ in the harmonic coordinate and $r$ in
Eq.~(\ref{eq:spherical ansatz}) is described by~\cite{Xu:2022frb}
\begin{equation}
  R''+\left(\nu'-\mu'+\frac{2}{r}\right)R'-\frac{2e^{2\mu}}{r^2}R=0\,.
\end{equation}
With the solution derived in Appendix~\ref{app:order solution}, one can obtain
the relation between $R$ and $r$ as 
\begin{equation}
  R=r-M+\frac{\xi}{2\kappa}\left(1-\frac{\xi}{2\kappa}\right)^2\frac{M^2}{2r}\left(\frac{Q}{M}
  \right)^4+\mathcal{O}\left(\frac{Q}{M}\right)^6\,.
  \label{eq:coordtransform}
\end{equation}
The leading order correction caused by the bumblebee charge in this relation is
at the order of $(Q/M)^4$, which will not enter the later calculations as the
leading order effects in light propagation and orbital motion are both at the
$(Q/M)^2$ order.

Solving Eqs.~(\ref{eq:light prop}--\ref{eq:eq of b}) to the leading order of
$Q/M$ and combining the coordinate transformation in
Eq.~(\ref{eq:coordtransform}), one can obtain $\Delta_{\rm prop}$ in the
harmonic coordinate
\begin{equation}
  \Delta_{\rm prop}=\Delta_{\rm R}+\Delta_{\rm S}+\Delta_{\rm Q}\,,
\end{equation}
where $\Delta_{\rm R}$ and $\Delta_{\rm S}$ are the so-called R{\"{o}}mer delay
and Shapiro delay respectively~\cite{Damour:1986}, given by
\begin{align}
  \Delta_{\rm R} &=-R\cos{\Delta\phi}\,,\ \\ 
  \Delta_{\rm S} &=-2M\ln(R+R\cos\Delta\phi)\,,
\end{align}
and $\Delta_{\rm Q}$ is the additional light propagation delay caused by the
bumblebee charge $Q$ at the leading order
\begin{equation}
  \Delta_{\rm
  Q}=-\frac{Q^2}{4R}\left(1-\frac{\xi}{2\kappa}\right)\left(\frac{3\Delta\phi}{
  \sin\Delta\phi}+\cos\Delta\phi\right)\,.
\end{equation}
Similarly to the leading order Shapiro delay, the leading order expression of
$\Delta_{\rm Q}$ diverges at the point $\Delta\phi=\pi$, as the leading order
calculation is effectively done by integrating along a straight line that will
cross the BH when $\Delta\phi=\pi$~\cite{Klioner:2010pu}. Here we only include
the leading order Shapiro delay, which is at 1\,PN order. As including 
higher-order terms is usually expected to further break the degeneracies between
parameters in such a problem, the treatment here is rather conservative. It is
sufficient for our purpose here and we leave the problem of constructing a more
delicate timing model for future studies.

The orbital motion of a pulsar with leading order corrections of $Q/M$ can be
calculated analytically using the so-called osculating orbital
elements~\cite{poisson_will_2014} or a similar procedure described by
\citet{Damour:1985}. However, those methods are hard to extend when considering 
higher order corrections or combining other possible effects such as the BH
spin. Simple analytic formula does not exist in general cases. Therefore, we
adopt the expansion form of the equations of motion in the harmonic coordinate
\begin{equation}
  \ddot{\bm{R}}=-\frac{M}{R^2}\hat{\bm{n}}-\frac{M}{R^2}\left[\left(-
  \frac{4M}{R}+v^2\right)\hat{\bm{n}}-4\dot{R}
  \,\bm{v}\right]+\left(1-\frac{\xi}{2\kappa}\right)
  \frac{Q^2}{R^3}\hat{\bm{n}}+\cdots\,,
\end{equation}
where $\bm{R}$ points from the central BH to the pulsar, the dots denote time
derivatives,  $\hat{\bm{n}}\equiv\bm{R}/R$, $\bm{v}\equiv\dot{\bm{R}}$, and
$v\equiv|\bm{v}|$. We only keep the 1\,PN term and the leading order correction
in $Q/M$ in this expansion for similar reasons that have been argued before.

The relation between the pulsar's proper time and the coordinate time has an
expansion form as 
\begin{equation}
  \frac{{\rm d}T}{{\rm
  d}t}=1-\frac{M}{R}-\frac{v^2}{2}+\frac{Q^2}{2R^2}\left(1-\frac{\xi}
  {2\kappa}\right)+\cdots\,.
\end{equation}
We integrate this equation to obtain the Einstein delay $\Delta_{\rm E}$. 

\begin{figure}[tbp]
  \centering 
  \includegraphics[width=14cm]{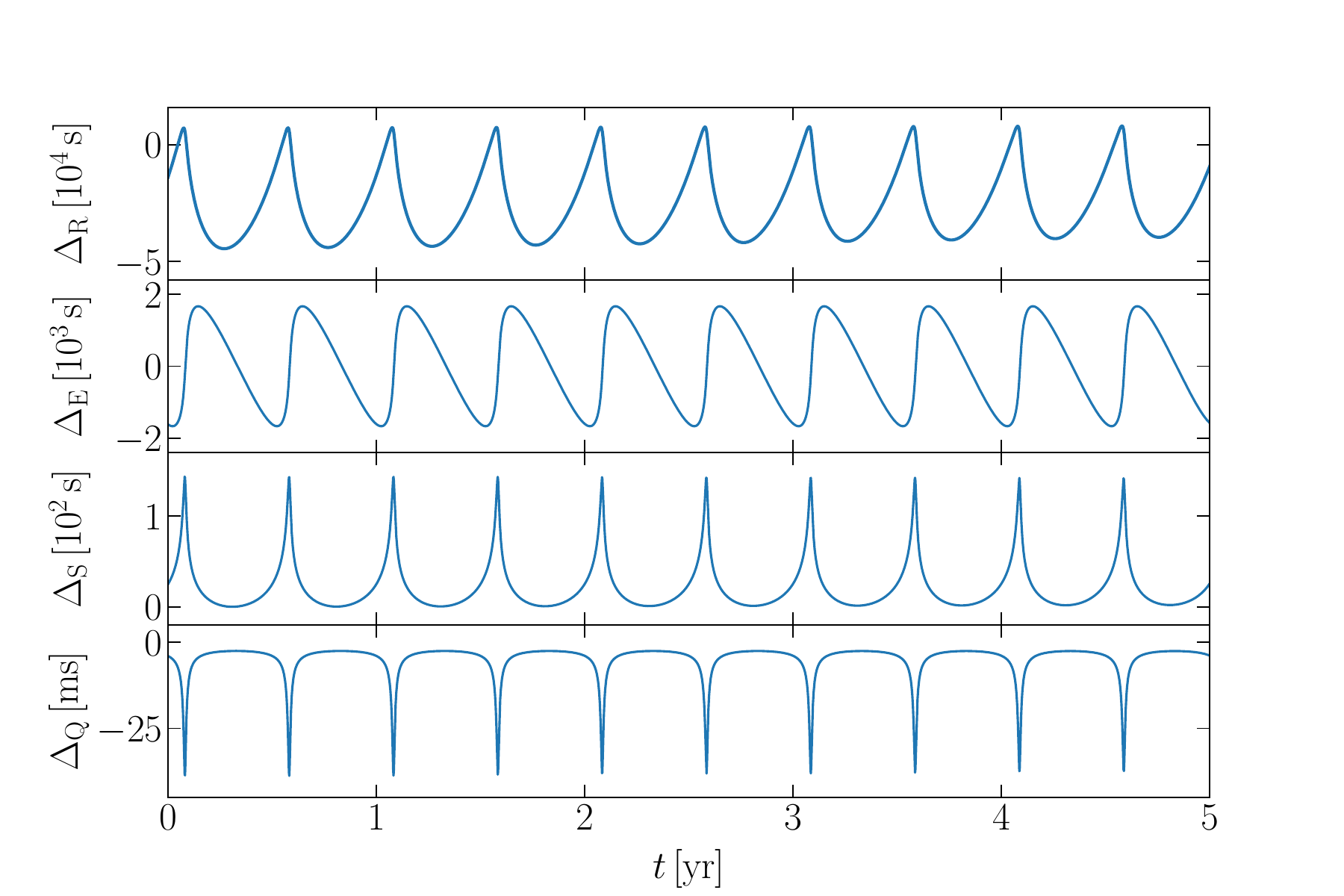}
  \caption{\label{fig:delays} Various time delays corresponding to the pulsar
  orbit shown in Fig.~\ref{fig:orbit} in the bumblebee gravity.}
\end{figure}

After dropping the constant term, one finally has 
\begin{equation}
	t^{\rm TOA}=T+\Delta_{\rm E}+\Delta_{\rm R}+
\Delta_{\rm S}+\Delta_{\rm Q} \,.
\end{equation}
In Fig.~\ref{fig:delays}, we show these time delays for the orbit illustrated
before with $\xi=\kappa$ and $Q=M$. The dominant modification from the bumblebee
charge is in the R\"omer delay at the 1\,PN level, due to the additional 
periastron advance in the bumblebee gravity. The additional propagation time
caused by the bumblebee charge has a similar behavior as the Shapiro delay but
with an opposite sign, and it is much smaller than the Shapiro delay.

\section{Parameter estimation}
\label{sec:esti}

As a standard procedure for parameter estimation in the pulsar timing 
process~\cite{Damour:1986,Hobbs:2006cd}, we use the covariance matrix,
\begin{equation}
  C_{\alpha\beta}=\left(\frac{\partial^2\mathcal{L}}{\partial\Theta^\alpha\partial\Theta^\beta}
  \right)^{-1}\,,
\end{equation}
to estimate the measurement uncertainties of the parameters
$\big\{\Theta^{\mu}\big\}$ in the timing model. The log-likelihood function
$\mathcal{L}$ is defined via~\cite{Damour:1986}
\begin{align}
  \mathcal{L} &=-\ln P \big(\Theta \big|t^{\rm TOA}\big)\,,\ \\ 
  P\big(\Theta\big|t^{\rm TOA}\big) &\propto\exp\left(-\frac{1}
  {2\nu_p^2}\sum_{i=1}^{N_{\rm
  TOA}}\frac{\left[N^{(i)}(\Theta)-N^{(i)}(\bar{\Theta})\right]^2} {\sigma_{\rm
  TOA}^2}\right)\,,
\end{align}
where $N_{\rm TOA}$ is the total number of observed TOAs, $\sigma_{\rm TOA}$ is
the timing precision,  $N^{(i)}(\Theta)$ is the pulsar rotation number of the
$i$-th TOA calculated from the timing model with parameters $\big\{\Theta^\mu
\big\}$, and $\bar{\Theta}$ represents the true values of those parameters. We
list all the model parameters here 
\begin{equation}
  \big\{\Theta^\mu
  \big\}=\Big\{N_0,\nu_p,\dot{\nu}_p,P_b,e,i,\omega,\sigma,M,(Q/M)^2\Big\}\,.
\end{equation}
As the dependence on $Q$ only appears in the quadratic form, we choose $(Q/M)^2$
as the model parameter.  Here we treat the coupling constant $\xi$ as a given
parameter that describes the gravity model and do not estimate it
simultaneously.  However, one would have the probability to constrain $\xi$ and
$Q$ simultaneously with high enough timing precision so that the next to leading
order effects can be observed. We assume the pulsar spin frequency
$\nu_p=2\,{\rm Hz}$, corresponding to a normal pulsar, and take the timing 
precision $\sigma_{\rm TOA}$ to be 1\,ms in our simulations, which is achievable
for future observations~\cite{Liu:2011ae,Bower:2018mta}. The total observation
time is assumed to be 5 years, and TOAs are extracted weekly and uniformly in
time, which gives $N_{\rm TOA}=260$. 

\begin{figure}[tbp]
  \centering 
  \includegraphics[width=14cm]{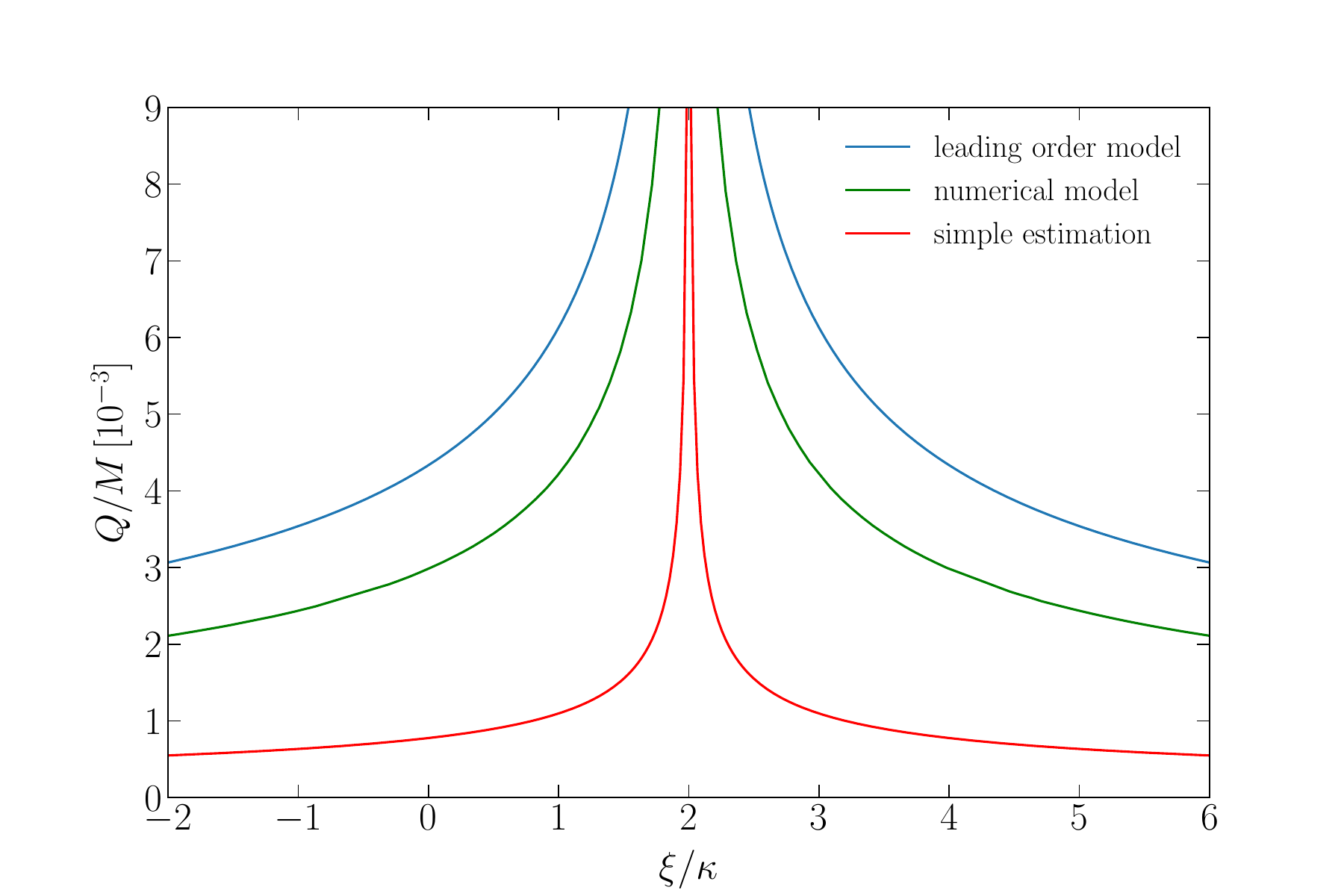}
  \caption{\label{fig:Q_xi} Constraints on $Q/M$ as functions of $\xi$. We
  compare the  results of a simple estimation and the two timing models (see
  text for more details). The pulsar orbital parameters are given in
  Eq.~(\ref{eq:orbit params}).}
\end{figure}

We have constructed a fully numerical timing model and a semi-analytic timing
model based on the leading order approximation in Sec.~\ref{sec:timing}. A
comparison of the estimation results of the two timing models are shown in
Fig.~\ref{fig:Q_xi}. For a pulsar orbit with orbital parameters shown in
Eq.~(\ref{eq:orbit params}), we present the constraint on $Q/M$ as a function of
$\xi$ obtained with the two timing models. In the simulation, we assume that GR
is correct, which means that the true value of $Q$ is set to be zero. The
constraint on $Q/M$ should be understood as an upper limit corresponding to the
square root of the uncertainty of $(Q/M)^2$. Intuitively, one can estimate upper
limits on $Q/M$ with $|k^{\rm Q}|\lesssim \delta k$, where $\delta k$ represents
the measurement precision of the periastron advance. For the assumed
pulsar-Sgr~A* system, $\delta k/k$ can be measured to a precision of $\sim
10^{-7}$~\cite{Liu:2011ae}. However, this kind of simple estimation is in 
general too optimistic because the degeneracy between the periastron advance
caused by the BH mass $M$ and the bumblebee charge $Q$ leads to larger
uncertainties for $M$ and $Q$ when doing a fully global determination of them.
Nevertheless, we still plot this simple estimation in Fig.~\ref{fig:Q_xi} as a
rough estimate. From this figure, one can see that the constraints from the
leading order model are slightly larger than the constraints from the numerical
model, with a factor of about 1.5. This is reasonable as we have ignored all
higher order effects of $Q/M$ in the leading order timing formula. Therefore, in
the following discussion we only present the estimation results obtained from
the leading order timing model.  One may expect that the results are similar if
the fully numerical model were adopted. One  also notices that the constraints
on $Q/M$ are weak when $\xi$ is around $2\kappa$ as the BH metric reduces to the
Schwarzchild solution when $\xi=2\kappa$~\cite{Xu:2022frb}.

\begin{figure}[tbp]
  \centering 
  \includegraphics[width=13cm]{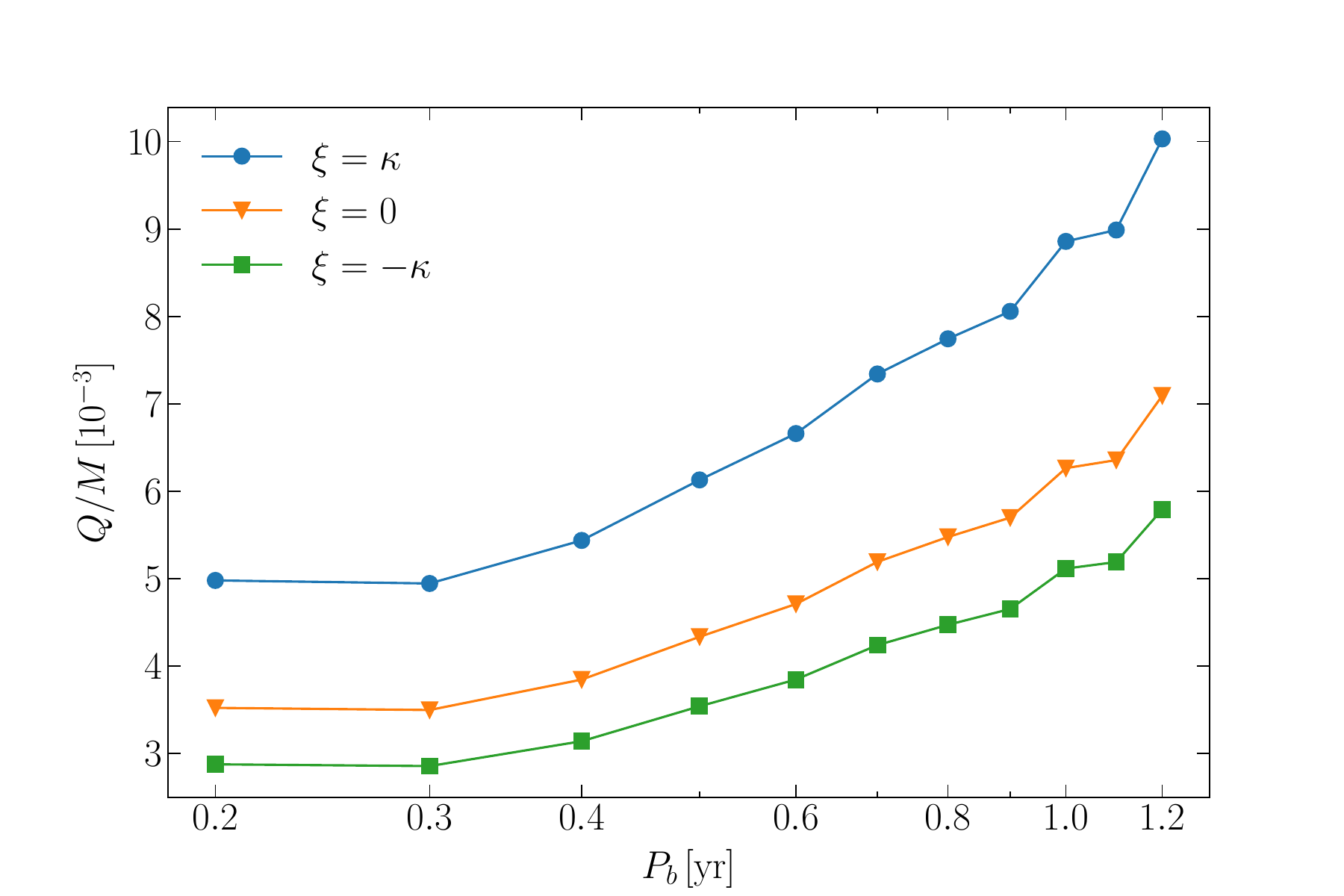}
  \caption{\label{fig:Q_Pb} Constraints on $Q/M$ as functions of the pulsar
  orbital period $P_b$ for $\xi=\kappa$, $\xi=0$ and $\xi=-\kappa$. Other
  orbital parameters are shown in Eq.~(\ref{eq:orbit params}).}
\end{figure}
\begin{figure}[tbp]
  \centering 
  \includegraphics[width=13cm]{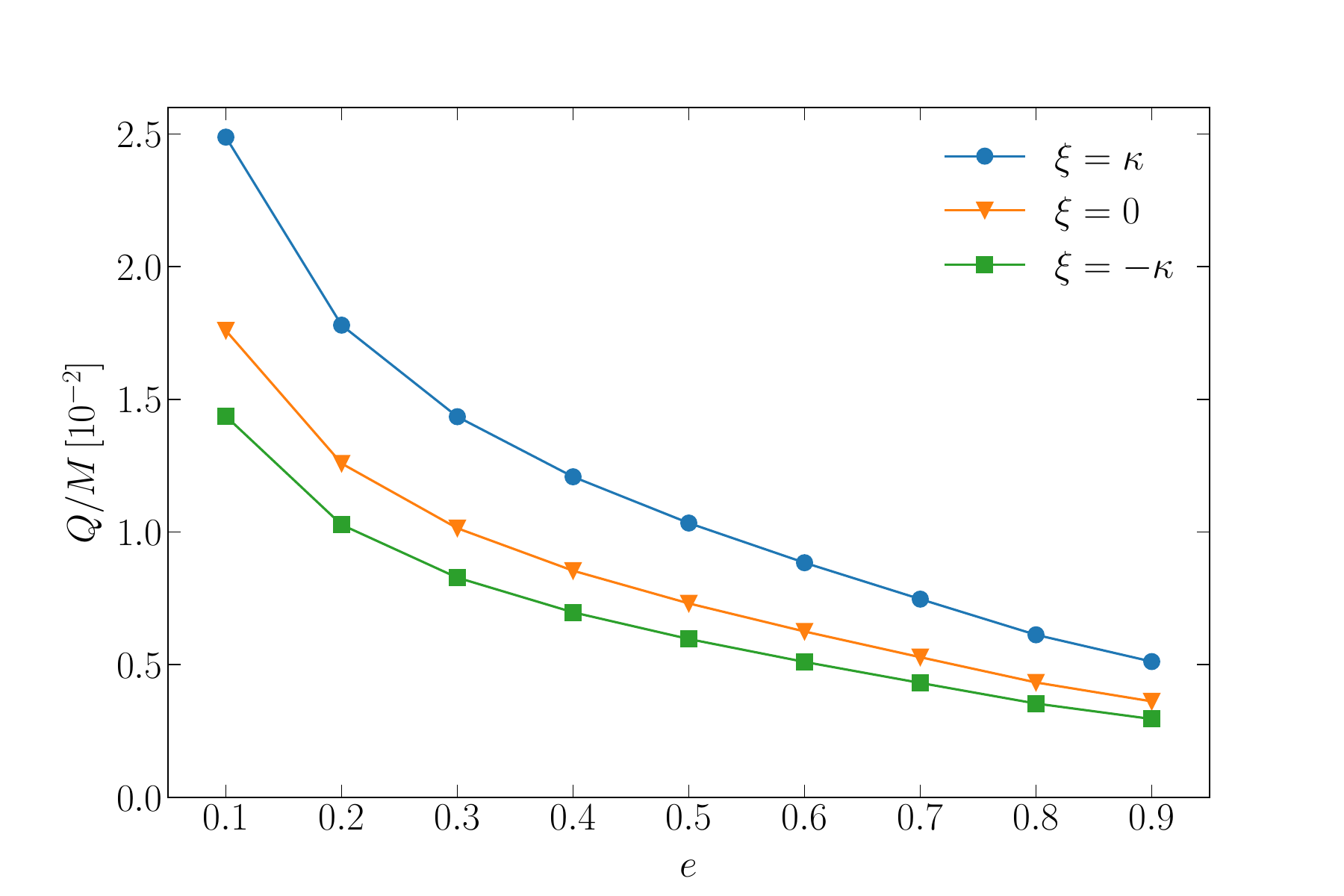}
  \caption{\label{fig:Q_e} Constraints on $Q/M$ as functions of the pulsar
  orbital eccentricity $e$ for $\xi=\kappa$, $\xi=0$ and $\xi=-\kappa$. Other
  orbital parameters are shown in Eq.~(\ref{eq:orbit params}).}
\end{figure}

In order to investigate more pulsar-Sgr\,A* systems than the prototype system in
Fig.~\ref{fig:Q_xi}, we vary the orbital parameters.  Figure~\ref{fig:Q_Pb}
shows the constraints on $Q/M$ as functions of the pulsar orbital period $P_b$
for $\xi=\kappa$, $\xi=0$ and $\xi=-\kappa$.  We can see an approximate symmetry
with respect to $\xi=2\kappa$ from the leading order metric solution.
Therefore, with the constraints on $Q/M$ being almost symmetric, we only show
the $\xi<2\kappa$ cases. The upper limits for $Q/M$ decrease when $P_b$ becomes
smaller as both the secular and periodic effects become larger when the pulsar
gets closer to the SMBH. We also investigate the influence of the orbital 
eccentricity and the results are shown in Fig.~\ref{fig:Q_e}. Compared to the
orbital period, a higher orbital eccentricity is relatively more important for
constraining the bumblebee charge.  This is because that the largest effect
caused by the bumblebee charge is the additional periastron advance, which
becomes degenerate with the orbital period when the orbital eccentricity is
small.  For a pulsar with an orbital period of 0.5~yr and an orbital
eccentricity of 0.8, one can expect to constrain $Q/M$ to $\sim
\mathcal{O}(10^{-3} \mbox{--} 10^{-2})$ when $|\xi/2\kappa-1|\gtrsim1$.

\section{Discussion}
\label{sec:discussion}

We have explored  pulsar-SMBH systems in the bumblebee gravity and discussed
possible constraints on the bumblebee charge of Sgr~A* by timing a pulsar
orbiting around it. We first constructed a timing model based on the fully
numerical calculations of the BH metric, light propagation, and pulsar orbital
motion in the bumblebee gravity. Considering the future timing precision for 
such GC pulsars and current constraints on the bumblebee charge of 
Sgr~A*~\cite{Xu:2022frb, Xu:2023xqh}, we also constructed a timing model only
including the leading order effects based on the expansion solution of the 
metric that is more efficient and flexible than the fully numerical approach.
Our simulation shows that for a pulsar with an orbital period $P_b\sim 0.5\,{\rm
yr}$, an orbital eccentricity $e\sim 0.8$, a 5-yr observation with weekly
recorded TOAs, and a timing precision of 1\,ms, one can constrain the bumblebee
charge of Sgr~A* to be $\sim \mathcal{O}(10^{-3} \mbox{--} 10^{-2})$ for
$|\xi/2\kappa-1|\gtrsim1$, which is much better than the current constraints
from the EHT observation~\cite{Xu:2023xqh}, and at the same order of magnitude
as the expected constraint from the extreme-mass-ratio inspiral observation with
the Laser Interferometer Space Antenna (LISA)~\cite{Liang:2022gdk}.

The limitation of the timing model built here is that we have only considered
the static spherical BHs in the bumblebee gravity. For spinning BHs, the
spin-orbit coupling and the quadruple moment of the BH both have detectable
effects in the pulsar timing observation~\cite{Wex:1998wt, Liu:2011ae,
Psaltis:2015uza, Hu:2023ubk}.  Also, the complex environment around Sgr~A* perturbs
the system~\cite{Merritt:2009ex, Hu:2023ubk}. The periastron advance caused by these
effects is degenerate with that caused by the bumblebee charge, which brings
larger uncertainties in the measurement even though one may separate these
effects by considering their periodic effects or combining other observations. 
Nevertheless, spinning BHs with bumblebee charges might introduce additional
features in TOAs that can be used to break the degeneracy.  To make a more
reliable estimation, one should extend the timing model to include, for example,
the spin of Sgr~A*. Although complete solutions of arbitrarily rotating BHs in
the bumblebee gravity have not been obtained yet, it is still possible to extend
the leading order timing model by adding the spin-orbit coupling term in the
equations of motion using some specific slowly-rotating BH solutions obtained
by~\citet{Ding:2020kfr}. We leave these kinds of extension for future studies.

\appendix
\section{An analytic solution in orders of $Q/M$}
\label{app:order solution}

To obtain a solution of Eqs.~(\ref{eq:field eqs1}--\ref{eq:field eqs3}) arranged
in the order of $Q/M$, one can start with the expansion
\begin{eqnarray}
  b_t &=& b_{t0}\epsilon+b_{t1}\epsilon^3+\mathcal{O}\left(\epsilon^5\right)\,,
  \label{eq:expand metric1}\\
  g_{tt} &=& g_{tt0}+g_{tt1}\epsilon^2+g_{tt2}\epsilon^4+
  \mathcal{O}\left(\epsilon^6\right)\,,\label{eq:expand metric2}\\
  g_{rr} &=& g_{rr0}+g_{rr1}\epsilon^2+g_{rr2}\epsilon^4+
  \mathcal{O}\left(\epsilon^6\right)\,,\label{eq:expand metric3}\\
  r_h &=& r_{h0}+r_{h1}\epsilon^2+r_{h2}\epsilon^4+
  \mathcal{O}\left(\epsilon^6\right)\,,
\end{eqnarray}
where $b_{t0}, b_{t1}, \cdots$ are functions of $r$ and $\epsilon=Q/M$ is  a
notation to denote the different orders of $Q/M$. Substituting these equations
into the field equations, one can obtain the equations for each order. The
expansion for the radius of event horizon $r_h$ is used when applying the
boundary conditions.

For a BH with mass $M$ and bumblebee charge $Q$, we choose the zeroth order
solution to be a Schwarzchild solution with mass $M$, i.e., 
\begin{align}
  g_{tt0} &=-1+\frac{2M}{r}\,,\\
  g_{rr0} &=\left(1-\frac{2M}{r}\right)^{-1}\,,\\ 
  r_{h0}&=2M\,.
\end{align}
Different choices of the zeroth order solution can change the higher order terms
but the entire series refer to the same exact solution.

From Eq.~(\ref{eq:field eqs1}), we obtain the ordinary differential equation for
$b_{t0}$, which reads as
\begin{equation}
  b_{t0}''+\frac{2}{r}b_{t0}'=0\,.
\end{equation}
The general solution of this equation, $b_{t0}=c_1+c_2/r$, has two integration
constants, which can be determined by the two boundary conditions 
\begin{align}
	b_t &=0 \,, &  \mbox{when} \quad r=r_h \\
	-r^2b_t'&\rightarrow \sqrt{2/\kappa}\,Q \,, & \mbox{when} \quad
	r\rightarrow\infty
\end{align}
Note that the boundary conditions should also be satisfied order by order. Thus
one can obtain 
\begin{equation}
  b_{t0}=-\frac{1}{\sqrt{2\kappa}}\left(1-\frac{2M}{r}\right)\,.
\end{equation}

Then from Eqs.~(\ref{eq:field eqs2}--\ref{eq:field eqs3}) one can derive the
ordinary differential equation for $g_{tt1}$, which reads as
\begin{equation}
  g_{tt1}''+\frac{2}{r}g_{tt1}'+\frac{2M^2}{r^4}\left(1-\frac{\xi}{2\kappa}\right)=0\,.
\end{equation}
The general solution is
\begin{equation}
  g_{tt1}=c_1+\frac{c_2}{r}-\frac{M^2}{r^2}\left(1-\frac{\xi}{2\kappa}\right)\,.
\end{equation}
The boundary conditions for $g_{tt1}$ come from the requirements that when
$r\rightarrow\infty$, $g_{tt}\rightarrow -1$ and $r^2\nu'\rightarrow M$. So one
finds $c_1=c_2=0$. After working out $g_{tt1}$, $g_{rr1}$ and $r_{h1}$ can be
determined directly from Eq.~(\ref{eq:field eqs1}) together with $g_{tt}(r_h)=0$
respectively.

Repeating the above procedure, one can obtain the solution for a BH with mass
$M$ and bumblebee charge $Q$ in orders of $Q/M$. The result up to the order of
$(Q/M)^6$ is listed here
\begin{align}
  b_t&=-\frac{1}{\sqrt{2\kappa}}\left(1-\frac{2M}{r}\right)\frac{Q}{M}\nonumber\\
  &\ \ \ \ -\frac{1}{4}\frac{1}
  {\sqrt{2\kappa}}\left(1-\frac{\xi}{2\kappa}\right)\left[\left(1-\frac{\xi}{2\kappa}\right)
  +\frac{4M^2}{r^2}\frac{\xi}{2\kappa}\right]\left(\frac{Q}{M}\right)^3
  +\mathcal{O}\left(\frac{Q}{M}\right)^5\,,\\
  g_{tt}&=-1+\frac{2M}{r}-\left(1-\frac{\xi}{2\kappa}\right)\frac{Q^2}{r^2}\nonumber\\
  &\ \ \ \ -\frac{\xi}{2\kappa}\left(
  1-\frac{\xi}{2\kappa}\right)^2\left(1-\frac{M}{r}\right)\frac{M^2}{r^2}\left(\frac{Q}{M}
  \right)^4+\mathcal{O}\left(\frac{Q}{M}\right)^6\,,\\
  g_{rr}&=\left[1-\frac{2M}{r}+\left(1-\frac{\xi}{2\kappa}\right)\frac{Q^2}{r^2}\right.
  \nonumber\\
  &\ \ \ \
  +\left.\frac{\xi}{2\kappa}\left(1-\frac{\xi}{2\kappa}\right)^2\left(2-\frac{3M}{r}\right)
  \frac{M^2}{r^2}\left(\frac{Q}{M}\right)^4\right]^{-1}+\mathcal{O}\left(\frac{Q}{M}\right)^6\,,\\
  \frac{r_h}{M}&=1+\sqrt{1-\left(1-\frac{\xi}{2\kappa}\right)\left(\frac{Q}{M}\right)^2}
  -\frac{1}{4}\frac{\xi}{2\kappa}\left(1-\frac{\xi}{2\kappa}\right)^2\left(\frac{Q}
  {M}\right)^4+\mathcal{O}\left(\frac{Q}{M}\right)^6\,.
\end{align}
The forms of $g_{rr}$ and $r_h$ are chosen such that the RN solution and the
Schwarzschild metric can be recovered by the leading order terms at $\xi=0$ and
$\xi=2\kappa$ respectively.

As discussed in Sec.~\ref{sec:BBBH}, the charged BHs in the bumblebee gravity
are indistinguishable with RN BHs in GR at the leading order of $Q/M$. However,
this degeneracy can be broken by considering higher order terms as the
combination of $\xi$ and $Q$ changes in the next-to-leading order contributions.

\acknowledgments

This work was supported by the National SKA Program of China (2020SKA0120300),  
the National Natural Science Foundation of China (11975027, 11991053, 12247128,
11721303), the China Postdoctoral Science Foundation (2021TQ0018), the Max
Planck Partner Group Program funded by the Max Planck Society, and the
High-Performance Computing Platform of Peking University.

\bibliographystyle{apsrev4-1}
\bibliography{refs}

\end{document}